\documentclass[aps, twocolumn, letterpaper, prl, showpacs]{revtex4}

\usepackage{amsmath}
\usepackage{amssymb}
\usepackage{mathrsfs}
\usepackage{xspace}
\usepackage{graphicx}

\usepackage{ifpdf}
\ifpdf
\fi

\newcommand{\eq}[1]{(\ref{#1})}
\newcommand{\Eq}[1]{Eq.~(\ref{#1})}
\newcommand{\Eqs}[1]{Eqs.~(\ref{#1})}
\newcommand{\Fig}[1]{Fig.~\ref{#1}}
\newcommand{\Ref}[1]{Ref.~\cite{#1}}
\newcommand{\Refs}[1]{Refs.~\cite{#1}}

\newcommand{\eg}{{e.g.,\/}\xspace}
\newcommand{\ie}{{i.e.,\/}\xspace}

\newcommand{\mc}[1]{\mathcal{#1}}
\newcommand{\mcc}[1]{\mathfrak{#1}}
\newcommand{\msf}[1]{\mathsf{#1}}
\newcommand{\mcu}[1]{\mathscr{#1}}
\renewcommand{\vec}[1]{{\boldsymbol{\rm #1}}}
\newcommand{\favr}[1]{\langle #1 \rangle}
\newcommand{\msection}[1]{\textit{#1.}\ ---\ }

\begin{document}

\title{Ponderomotive forces \textit{on} waves in modulated media}

\author{I.~Y. Dodin and N.~J. Fisch}
\affiliation{Princeton Plasma Physics Laboratory, Princeton University, Princeton, New Jersey 08543, USA}

\begin{abstract}
Nonlinear interactions of waves via instantaneous cross-phase modulation can be cast in the same way as ponderomotive wave-particle interactions in high-frequency electromagnetic field. The ponderomotive effect arises when rays of a probe wave scatter off perturbations of the underlying medium produced by a second, modulation wave, much like charged particles scatter off a quasiperiodic field. Parallels with the point-particle dynamics, which itself is generalized by this theory, lead to new methods of wave manipulation, including asymmetric barriers for light.
\end{abstract}



\maketitle

\bibliographystyle{brief}

\msection{Introduction} One of the curious effects in wave-particle interactions is that a rapidly oscillating electromagnetic (EM) field can produce a time-averaged force, known as the ponderomotive force, on any particle that is charged or, more generally, has a nonzero polarizability \cite{foot:pp, foot:minogin}. This effect, which can be attractive or repulsive depending on a specific interaction, is widely used in various applications ranging from atomic cooling to plasma confinement \cite{foot:nobel, foot:invited}. Moreover, it was shown recently that ponderomotive forces can cause nonreciprocal dynamics, such as one-way-wall effects \cite{foot:invited, foot:cd, my:ratchet, foot:atom}, and perform other nonintuitive transformations of the particle phase space \cite{foot:nonadiab}. As it turns out, and as we argue in this paper, the same effects can be practiced also \textit{on waves} (including EM, acoustic, or, for that matter, any propagating signals), if the parameters of the underlying medium are modulated quasiperiodically in time or space. 

Specifically, what we show here is that the interaction of waves in Kerr media via cross-phase modulation (XPM) can be cast in the same way as ponderomotive wave-particle interactions. The ponderomotive effect arises when rays of a geometrical-optics (GO) probe wave (PW) scatter off perturbations of the underlying medium produced by a second, modulation wave (MW), much like particles scatter off a quasiperiodic EM field. In contrast to the PW refraction caused by gradual changes of the medium average parameters (``slow'' nonlinearity), the ponderomotive effect on rays is instantaneous and can be inferred from the PW linear dispersion alone, irrespective of the medium evolution. 

The practical utility of this finding is threefold: (i)~Based on parallels with the wave-particle dynamics, new qualitative effects for wave-wave interactions are predicted. Examples of such effects that we put forth here are ponderomotive reflection (which must not to confused with resonant, Bragg reflection) and asymmetric barriers for light. (ii)~The XPM via instantaneous nonlinearities can now be described, both generally and quantitatively, beyond the special cases studied in literature \cite{foot:xpm, foot:mendonca, foot:tsytovich}. In particular, we derive equations for the PW continuous ponderomotive dynamics that remain manifestly conservative even when the medium average parameters slowly evolve in time or space. (iii)~The traditional theory of ponderomotive forces on point particles, which happens to be subsumed under the new formulation, is also generalized now, specifically to quasiclassical interactions. 

\msection{Basic equations} Consider a linear PW propagating in a general dissipationless medium such that the GO approximation is justified. This implies, in particular, that the wave resides on a single branch of the dispersion relation, even though parameters of the medium may vary with time $t$ and (arbitrarily curved) spatial coordinates $\vec{x}$. Assuming, for simplicity, that the wave is of the scalar type (which includes vector waves with fixed polarization too), it can hence be assigned a single canonical phase $\theta(t, \vec{x})$, a scalar action density $\mc{I}(t, \vec{x})$, and the Lagrangian density~\cite{ref:hayes73, foot:wkin}
\begin{gather}\label{eq:L}
\mcc{L} = -\mc{I} \big[\partial_t \theta + \omega(t, \vec{x}, \nabla \theta)\big].
\end{gather}
Both $\theta(t, \vec{x})$ and $\mc{I}(t, \vec{x})$ are independent functions here, so \Eq{eq:L} generates two Euler-Lagrange equations,
\begin{gather}
\partial_t \theta + \omega = 0,\label{eq:tJ}\\
\partial_t \mc{I} + \nabla \cdot (\mc{I}\vec{v}_{\rm g}) = 0,\label{eq:act}
\end{gather}
where $\vec{v}_{\rm g}(t, \vec{x}) \doteq \partial_{\vec{k}}\omega(t, \vec{x}, \vec{k}(t, \vec{x}))$ is the group velocity, and $\vec{k} \doteq \nabla \theta$ is the local wave number. (We use the symbol $\doteq$ to denote definitions.) Equation \eq{eq:tJ} is of the Hamilton-Jacobi type and serves as the dispersion relation, $\omega = \omega(t, \vec{x}, \vec{k})$, since $-\partial_t \theta$ is, by definition, the wave local frequency. Equation \eq{eq:act} has the form of a continuity equation and represents the action conservation theorem. To close this set of equations, the so-called consistency relations are used,
\begin{gather}\label{eq:cons}
\partial_t \vec{k} + \nabla \omega = 0, \quad \nabla \times \vec{k} = 0,
\end{gather}
which flow from the definitions of $\omega$ and $\vec{k}$. Equations \eq{eq:tJ}-\eq{eq:cons} are also known as the Whitham equations and subsume the familiar ray equations~\cite{foot:stix},
\begin{gather}\label{eq:rays0}
\dot{x}^\ell = \partial \omega(t, \vec{x}, \vec{k})/\partial k_\ell, \quad \dot{k}_\ell = - \partial \omega(t, \vec{x}, \vec{k})/\partial x^\ell,
\end{gather}
as their characteristics~\cite{foot:whitham, my:amc}. 

\msection{Reduced equations} Suppose now that $\omega = \bar{\omega} + \tilde{\omega}$, where $\tilde{\omega}(t, \vec{x}, \vec{k}) = \text{Re}\,[\tilde{\omega}_c(t, \vec{x}, \vec{k})\, e^{i\Theta(t, \vec{x})}]$ is a small perturbation. We term the latter a MW and introduce its frequency $\Omega \doteq - \partial_t \Theta$ and wave vector $\vec{K} \doteq \nabla \Theta$. Suppose also that $\Theta$ evolves slowly enough, so that the GO approximation for the PW holds (and, in particular, resonant effects like Bragg scattering do not occur). On the other hand, we will assume that $\Theta$ evolves fast compared to the rate at which $\bar{\omega}$ and the MW parameters ($\Omega$, $\vec{K}$, and the amplitude) change in time and space. Hence we can unambiguously introduce the slow, $\Theta$-independent, or adiabatic dynamics, which is done as follows. 

Let us express the PW phase as $\theta = \bar{\theta} + \tilde{\theta}$ and the PW action density as $\mc{I} = \bar{\mc{I}} + \tilde{\mc{I}}$, where $\tilde{\theta}$ and $\tilde{\mc{I}}$ are oscillating functions of the order of $\tilde{\omega}_c$; also, $\bar{\theta} \doteq \favr{\theta}$, and $\bar{\mc{I}} \doteq \favr{\mc{I}}$, where the angular brackets denote local averaging over $\Theta$. As usual \cite{foot:bgk, foot:whitham}, the Lagrangian density of slow, adiabatic dynamics can then be calculated as $\mcu{L} = \favr{\mcc{L}}$. After neglecting terms of order $|\tilde{\omega}_c|^r$ with $r > 2$, one gets 
\begin{gather}\label{eq:L2}
\mcu{L} = - \bar{\mc{I}} \big[\partial_t\bar{\theta} + w(t, \vec{x}, \nabla \bar{\theta})\big],
\end{gather}
where $w(t, \vec{x}, \bar{\vec{k}}) = \bar{\omega} + \favr{\nabla\tilde{\theta} \cdot \bar{\omega}_{\bar{\vec{k}}\bar{\vec{k}}} \cdot \nabla\tilde{\theta}}/2 + \favr{\tilde{\omega}_{\bar{\vec{k}}} \cdot \nabla\tilde{\theta}} + h$, and $h \doteq \favr{(\partial_t \tilde{\theta} + \bar{\omega}_{\bar{\vec{k}}} \cdot \nabla\tilde{\theta} + \tilde{\omega})\,\tilde{\mc{I}}}$. Both $\bar{\omega}$ and $\tilde{\omega}$ are evaluated here at $(t, \vec{x}, \bar{\vec{k}})$, and the index $\bar{\vec{k}}$ denotes the corresponding partial derivative. The quiver phase, $\tilde{\theta}$, satisfies the linearized equation $\partial_t \tilde{\theta} + \bar{\omega}_{\bar{\vec{k}}} \cdot \nabla \tilde{\theta} + \tilde{\omega} = 0$, with $\bar{\vec{k}} = \nabla \bar{\theta}$. This leads to $h = 0$ and also $\tilde{\theta}= - i\tilde{\omega}/(\Omega - \vec{K} \cdot \bar{\omega}_{\bar{\vec{k}}})$. A straightforward calculation then yields
\begin{gather}\label{eq:H}
w(t, \vec{x}, \bar{\vec{k}}) = \bar{\omega} + \frac{\vec{K}}{4} \cdot
\frac{\partial}{\partial \bar{\vec{k}}}\left(
\frac{|\tilde{\omega}_c|^2}{\Omega - \vec{K} \cdot \bar{\omega}_{\bar{\vec{k}}}}
\right),
\end{gather}
where, within the adopted accuracy, $\bar{\omega}_{\bar{\vec{k}}}$ can be replaced with $\bar{\vec{v}}_{\rm g} \doteq \partial_{\bar{\vec{k}}} \omega_0$, and $\omega_0$ is the unperturbed PW frequency evaluated at $\bar{\vec{k}}$. (The possible difference between $\omega_0$ and $\bar{\omega}$ will become clear from examples below.) Equations \eq{eq:L2} and \eq{eq:H} also lead to dynamic equations akin to the original Whitham equations \eq{eq:tJ}-\eq{eq:cons}:
\begin{gather}
\partial_t \bar{\theta} + w = 0,\label{eq:bth}\\ 
\partial_t \bar{\mc{I}} + \nabla \cdot (\bar{\mc{I}}{\vec{u}}_{\rm g}) = 0,\label{eq:tJ2}\\
\partial_t \bar{\vec{k}} + \nabla \bar{\omega} = 0, \quad \nabla \times \bar{\vec{k}} = 0,\label{eq:cons2}
\end{gather}
where ${\vec{u}}_{\rm g} \doteq \partial_{\bar{\vec{k}}} w$. Then the corresponding ``oscillation-center'' (OC) ray equations, which can be considered as time-averaged \Eqs{eq:rays0}, are
\begin{gather}\label{eq:rays}
\dot{\bar{x}}^\ell = \partial w(t, \bar{\vec{x}}, \bar{\vec{k}})/\partial \bar{k}_\ell, 
\quad 
\dot{\bar{k}}_\ell = - \partial w(t, \bar{\vec{x}}, \bar{\vec{k}})/\partial \bar{x}^\ell.
\end{gather}
Here $w$ acts as the OC Hamiltonian of PW rays, or their ponderomotive Hamiltonian, so one may recognize \Eq{eq:H} as an extension to continuous waves of what is a known theorem in classical mechanics of discrete systems \cite{foot:madey, foot:kchi}. [The cause of this parallel is that \Eq{eq:tJ}, which describes the dispersion relation of a continuous wave, is identical to the Hamilton-Jacobi equation for a ray as a discrete quasiparticle governed by \Eqs{eq:rays0}.] From the particle analogy (cf., \eg \Ref{foot:cd}), one also obtains the adiabaticity condition underlying \Eqs{eq:L2}-\eq{eq:rays}; namely, in addition to the smallness of $\tilde{\omega}$, one must have
\begin{gather}\label{eq:tau}
\dot{\tau} \ll 1, \quad \tau \doteq |\Omega - \vec{K} \cdot \bar{\vec{v}}_{\rm g}|^{-1},
\end{gather}
where $\tau$ is the modulation time scale in the ray reference frame, and the time derivative is taken along rays.

Equations \eq{eq:L2}-\eq{eq:tau} are the main analytical results of our paper. They provide a new, \textit{general} description of the MW effect on the GO propagation of a nondissipative continuous PW in any medium with a Kerr-type, cubic nonlinearity \cite{foot:cubic}. (XPM via second-order nonlinearities does not appear in our picture because the Pockels effect, such as in \Ref{foot:pockels}, requires $\dot{\tau} \gtrsim 1$ and otherwise averages to zero.) Slow nonlinearities enter here through the dependence of $w$ on $\Theta$-averaged parameters of the medium. To assess this effect quantitatively, one merely needs to add the OC Lagrangian density of the medium to $\mcu{L}$ \cite{foot:pf, my:amc, foot:var} and calculate the medium evolution in response to the ponderomotive force that a MW produces \textit{on matter}. However, below we will focus instead on the general nonlinearity that is independent of the medium inertia. It can be viewed as an instantaneous ponderomotive effect that the MW produces \textit{directly on PW rays} and hence is termed ``ponderomotive refraction''.

\msection{Ponderomotive refraction} Even at small $\tilde{\omega}$, ponderomotive refraction can be a significant factor in the PW evolution, especially when the underlying medium is homogeneous and stationary. The effect can be particularly strong near the group-velocity resonance (GVR), $\Omega \simeq \vec{K} \cdot \bar{\vec{v}}_{\rm g}$. This is naturally understood for broad-spectrum PW pulses, as then the GVR can be (at least loosely) interpreted as the Cherenkov resonance between PW ``quanta'' and the MW \cite{foot:mendonca, foot:tsytovich, ref:bu14}. However, as seen from our theory, the GVR remains a peculiar regime even for homogeneous waves, in which case $\bar{\vec{v}}_{\rm g}$ does not have a transparent meaning of the envelope velocity. 

What is also remarkable is that \Eq{eq:H} describes ponderomotive refraction solely from the PW linear dispersion, without consideration of details of the nonlinear dynamics of the medium, in contrast to traditional theories \cite{foot:suscp}. Here are some examples. First, suppose a sound-like wave, $\omega(t, \vec{x}, \vec{k}) = kC(t, \vec{x})$, where $C(t, \vec{x}) = C_0(t, \vec{x}) + \text{Re}\,[\tilde{C}(t, \vec{x})\,e^{i\Theta(t, \vec{x})}]$, such that $C_0$ and $\tilde{C}$ are slow functions. Then $\bar{\omega} = \omega_0 = \bar{k} C_0$, and $\tilde{\omega} = \bar{k}\tilde{C}$, so \Eq{eq:H} yields
\begin{gather}\label{eq:ws}
w = \omega_0 \left[1 + \frac{\varepsilon^2\cos \chi}{U - \cos \chi} 
+ \frac{\varepsilon^2 \sin^2 \chi}{2(U - \cos \chi)^2}\right],
\end{gather}
where $U \doteq \Omega/(KC_0)$, $\varepsilon^2 \doteq |\tilde{C}|^2/(2C_0^2)$, and $\chi$ is the angle between $\bar{\vec{k}}$ and $\vec{K}$. Suppose, for simplicity, that $U \sim 1$ and that any spatial gradients are along $\vec{K}$, so the transverse wave vector, $\bar{\vec{k}}_\perp$, is conserved. At quasi-parallel propagation ($\bar{k}_\lVert \gg \bar{k}_\perp$, where $\bar{k}_\lVert$ is the parallel component of the wave vector), one then gets $w \simeq \bar{k}_\lVert C_{\rm eff}$, where $C_{\rm eff} \doteq C_0 [1 + \varepsilon^2/(U-1)]$; \ie the ponderomotive effect simply changes the sound speed from $C_0$ to $C_{\rm eff}$. In contrast, at quasi-transverse propagation ($\bar{k}_\lVert \ll \bar{k}_\perp$), one gets $w/(\bar{k}_x C_0) \simeq 1 + (p - \alpha)^2/2 + \phi$, where $p \doteq \bar{k}_\lVert/\bar{k}_\perp$, $\alpha \doteq - \varepsilon^2(1 + U^2)/U^3$, and $\phi \doteq \varepsilon^2/(2U^2)$. In particular, if $C_0$ is a constant, and $\alpha$ is time-independent, the latter can be removed by gauge transformation as an effective vector potential. Then the ponderomotive effect consists of ray repulsion by the effective scalar potential $\phi$. (As a side remark, we note that these effects are not captured by the standard, linear theory of mode conversion \cite{foot:mcop}, as the latter does not account for the dependence of ray scattering on the MW amplitude.)

\begin{figure}[b]
\centering
\includegraphics[width=.48\textwidth]{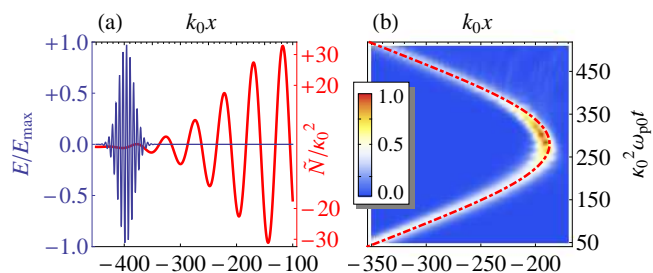}
\caption{(color online) Results of one-dimensional full-wave simulations of the EM pulse scattering in plasma off a density wave with stationary envelope, $\Omega = \omega_{p0}\kappa_0^2$, and $K = - 0.12k_0$; here $\kappa_0 \doteq ck_0/\omega_{p0} \lll 1$, and $k_0$ is the initial wave number. Slow nonlinearities are ignored, and oscillations at the constant carrier frequency $\omega_{p0}$ are mapped out. The electric field envelope $E$ is given in units of its maximum amplitude; $\tilde{N}$ is given in units $\kappa_0^2$; $t$ is given in units $\kappa_0^{-2}\omega_{p0}^{-1}$; $x$ is given in units $k_0^{-1}$. (a) Initial setup ($t = 0$); shown are $E$ (blue, narrow envelopes) and $\tilde{N}$ (red, wide envelope). (b)~$|E(t, x)|^2$ and the ray trajectory found by numerical integration of \Eqs{eq:rays}.}
\label{fig:barrier}
\end{figure}

As another example, consider an EM wave in nonmagnetized plasma with electron-density relative perturbation $\tilde{N}(t, \vec{x})$. Then $\omega(t, \vec{x}, \vec{k}) = [\omega_p^2(t, \vec{x}) + k^2 c^2]^{1/2}$, where $\omega_p = \omega_{p0}(1 + \tilde{N})^{1/2}$ is the plasma frequency, $\omega_{p0}$ is its unperturbed value, and $c$ is the speed of light. (Relativistic effects such as in \Ref{foot:ren} are neglected in this model.) Hence $\tilde{\omega} \simeq \tilde{N}\omega_{p0}^2/(2\omega_0)$, $\bar{\omega} \simeq (1 - \varepsilon^2)\omega_0$, and $\bar{\vec{v}}_{\rm g} = c^2\bar{\vec{k}}/\omega_0$, where $\omega_0 = (\omega_{p0}^2 + \bar{k}^2 c^2)^{1/2}$, $\varepsilon \doteq \tilde{N}_m \omega_{p0}^2/(4\omega_0^2)$, and $\tilde{N}_m$ is the amplitude of~$\tilde{N}$. One then~gets
\begin{gather}\label{eq:lp}
w = \omega_0 \big[1 - \varepsilon^2(\Omega^2 - K^2 c^2)/(\Omega - \vec{K} \cdot \bar{\vec{v}}_{\rm g})^2\big].
\end{gather}
Note that EM wave propagation in static modulated media, like photonic crystals \cite{foot:pc}, are described by \Eq{eq:lp} as a special case corresponding to $\Omega = 0$ (cf., \eg \Ref{ref:russel99}); then $w = \omega_0\,[1 + (\varepsilon/\msf{n}_\lVert)^2]$, where $\msf{n}_\lVert \equiv \bar{k}_\lVert c/\omega_0$ is the PW refraction index along $\vec{K}$. (Keep in mind, however, that this result applies only at large enough $\msf{n}_\lVert$, such that ${\vec{u}}_{\rm g}$ does not deviate much from $\bar{\vec{v}}_{\rm g}$.) Also let us consider the opposite limit, $\Omega \gg \vec{K} \cdot \bar{\vec{v}}_{\rm g}$. Assuming, for simplicity, that $\bar{k}$ is small enough, in this case one gets $w/\omega_{p0} \simeq 1 + c^2\bar{k}^2/(2\omega_{p0}^2) + \phi$, where $\phi = \varepsilon^2(\msf{N}^2 - 1)$ acts as an effective potential. Its sign is determined by the MW refraction index, $\msf{N} \doteq K c/\Omega$, which, in principle, can have any value, especially if the MW is produced by driven fields. Such $\phi$ thereby attracts PW rays if $\msf{N} < 1$ and repels them \cite{foot:slow} if $\msf{N} > 1$ (\Fig{fig:barrier}). (In particular, the latter case is realized when MW is one of the natural plasma waves, \eg an ion acoustic or Langmuir wave.) 

\begin{figure}
\centering
\includegraphics[width=.48\textwidth]{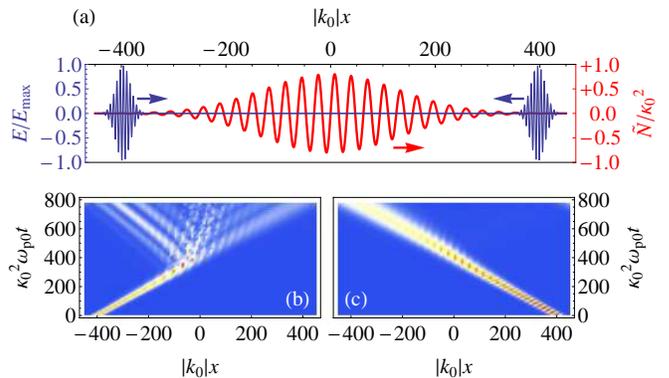}
\caption{(color online) Same as in \Fig{fig:barrier} but for $\Omega = 0.03\omega_{p0}\kappa_0^2$ and $K = 0.2|k_0|$. Two cases are considered, $k_0 > 0$ and $k_0 < 0$, with identical $|k_0|$. (a) Initial setup ($t = 0$); the arrows denote the directions of propagation. (b)~$|E(t, x)|$ for $k_0 > 0$. The pulse hits the GVR at about $k_0 x \simeq -100$; then it is partially transmitted and partially reflected, much like it is described for particles in \Ref{foot:bruhwiler}. (c)~$|E(t, x)|$ for $k_0 < 0$. As the signs of $k_0$ and $K$ are different, the pulse never hits the GVR and is fully transmitted.}
\label{fig:sim}
\end{figure}

\begin{figure}
\centering
\includegraphics[width=.48\textwidth]{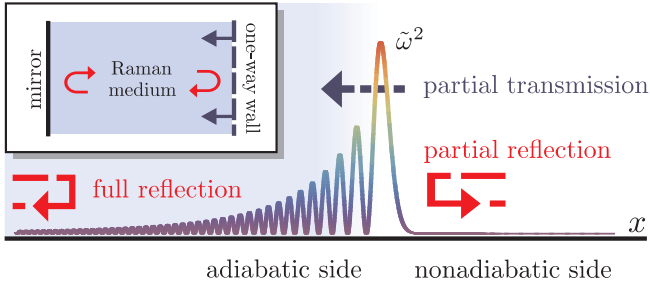}
\caption{(color online) Main figure: schematic of a one-dimensional asymmetric ponderomotive wall for PW rays. Rainbow-colored and oscillating is $\tilde{\omega}^2$, which determines $w$ [\Eq{eq:H}]. On the left, adiabatic side (shaded), the ponderomotive force reflects all rays below a certain frequency before they even reach the nonadiabatic region. On the right side, rays enter the nonadiabatic region first; then, depending on the initial phase, some of them are reflected, but others are transmitted. The feasibility of this type of asymmetric walls for charged particles was studied in detail, both analytically and numerically, in \Ref{my:ratchet}. Inset: possible scheme of a light concentrator (see the main text).}
\label{fig:onewaywall}
\end{figure}

Such an effective potential is similar to the adiabatic ponderomotive potential seen by point charges in a high-frequency EM field \cite{foot:pp}. It is hence also natural to extrapolate the wave-particle analogy to the Pockels regime ($\tau \gtrsim 1$), where the interaction is nonadiabatic \cite{foot:invited, foot:nonadiab}. Based on what is known about the particle dynamics in nonadiabatic ponderomotive barriers \cite{foot:invited, foot:nonadiab, foot:bruhwiler}, one readily anticipates that regions of strong MW can be arranged in this regime to scatter PW rays probabilistically and, when $\Omega$ is nonzero, also \textit{asymmetrically}. This is confirmed in simulations already for simple MW envelopes (\Fig{fig:sim}), and asymmetry can be made even stronger if the MW shape is specially adjusted (\Fig{fig:onewaywall}). (Notably, these manipulations are somewhat akin to those practiced via effective gauge fields on PWs in externally-driven lattices of multimode resonators \cite{foot:fang13b}. The difference is, however, that our ponderomotive forces exist in much simpler, continuous media and can be applied to single-mode pulses.) 

Such ``one-way walls'' can be used to direct rays in a ratchet manner, as suggested in \Refs{foot:cd, my:ratchet} for charged particles, or even to concentrate them, as proposed and implemented in \Ref{foot:atom} for atoms. For example, suppose a barrier shown in \Fig{fig:onewaywall} and the concentrator scheme as in the inset. With the aid of an additional mirror, this barrier can confine photons on its left side. That applies, of course, only for photons with energies below a certain threshold, whereas those transmitted from the right have energies above that threshold and thus can escape. (This is because, in the adiabatic domain, the motion is reversible, so any photon that once was at the top of the barrier can return there in the future.) However, like in the case of charged particles \cite{foot:cd, my:ratchet} and atoms \cite{foot:atom}, the barrier can serve as a one-way wall if dissipation is added. Suppose that a transmitted photon, with some frequency $\omega_1$, undergoes Raman decay into some natural oscillations with frequency $\omega_R$ and another photon with frequency $\omega_2 < \omega_1$. The former will dissipate, but, assuming $\omega_R \ll \omega_1$, this energy loss can be negligible. The second photon, however, now has a smaller (ideally, zero) probability to escape due to its lower energy and thus is stuck between the one-way wall and the mirror until it decays through the Raman cascade. Hence the photon density in that region will be higher than outside.

\msection{Classical particles as PWs} Although the above discussion appeals to understanding waves as particles, the particle dynamics itself can be viewed merely as a special case of the ponderomotive-refraction theory. To show this, we approach it quantum-mechanically as follows. Consider the Lagrangian density of a (scalar) quantum particle, $\mcc{L} = (i\hbar/2)(\psi^*\partial_t\psi - \psi\,\partial_t \psi^*) - \psi^* H(t, \vec{x}, - i\hbar\nabla)\psi$, where $H$ is a Hamiltonian, and $\psi(t, \vec{x})$ is the wave function in the spatial representation \cite{foot:wkin}. Let us represent this function through its (real) phase $S/\hbar$ and amplitude $\sqrt{\mc{I}}$. Specifically, we write $\psi = e^{iS(t, \vec{x})/\hbar}\sqrt{\mc{I}(t, \vec{x})}$, where $\mc{I}$ is now chosen to have units of number density rather than of action density, as before. Assuming that $\psi(t, \vec{x})$ is quasiclassical, we have $H(t, \vec{x}, - i\hbar\nabla)\psi \simeq H(t, \vec{x}, \nabla S)\psi$, so $\mcc{L}$ acquires the same form as in \Eq{eq:L}, 
\begin{gather}
\mcc{L} = - \mc{I} \big[\partial_t S + H(t, \vec{x}, \nabla S)\big], \label{eq:lqc}
\end{gather}
which does not contain $\hbar$. [Notably, and naturally, \Eq{eq:lqc} reproduces the well-known Lagrangian density of cold classical fluid \cite{foot:seliger} as a special case.] Therefore, if the particle Hamiltonian consists of a slow and rapidly-oscillating parts, $H = \bar{H} + \tilde{H}$, we can introduce a ponderomotive Lagrangian that describes the particle dynamics averaged over the oscillations of $\tilde{H}$:
\begin{gather}
\mcu{L} = -\bar{\mc{I}} \big[\partial_t \bar{S} + \mc{H}(t, \vec{x}, \nabla \bar{S})\big],\label{eq:aux100}\\
\mc{H}(t, \vec{x}, \vec{P}) = \bar{H} + \frac{\vec{K}}{4} \cdot
\frac{\partial}{\partial \vec{P}}\left(
\frac{|\tilde{H}|^2}{\Omega - \vec{K} \cdot \vec{V}}
\right).\label{eq:Heff}
\end{gather}
Here $\vec{V} \equiv \bar{\vec{v}}_{\rm g}$ is the OC velocity, and the remaining notation is self-explanatory. The model of \textit{point} particles, if needed, corresponds to $\mc{I}(t, \vec{x}) = \delta(\vec{x} - \vec{X}(t))\,|\text{det}\,\hat{g}|^{-1/2}$, where $\hat{g}$ is the spatial metric. The OC total Lagrangian, $\mc{L}$, is then obtained by integrating $\mcu{L}$ over the volume; that yields $\mc{L} = \vec{P} \cdot \dot{\vec{X}} - \mc{H}(t, \vec{X}, \vec{P})$ with $\vec{P} \doteq \nabla \bar{S}$, so $\mc{H}$ serves as the Hamiltonian for the canonical pair $(\vec{X}, \vec{P})$.

In particular, for an elementary particle with mass $m$ and charge $e$, one has $H(t, \vec{x}, \vec{P}) = \{m^2c^4 + [\vec{P} - e\vec{A}(t, \vec{x})/c]^2\}^{1/2} + e \varphi(t, \vec{x})$. Here $\vec{A} = \bar{\vec{A}} + \tilde{\vec{A}} $ and $\varphi = \bar{\varphi} + \tilde{\varphi}$ are the vector and scalar potentials; $\bar{\vec{A}}$ and $\bar{\varphi}$ describe quasistatic fields, if any; $\tilde{\vec{A}} = \text{Re}\,[\tilde{\vec{A}}_c\, e^{i\Theta(t, \vec{x})}]$ and $\tilde{\varphi} = \text{Re}\,[\tilde{\varphi}_c\, e^{i\Theta(t, \vec{x})}]$ describe a MW, which is now comprised of oscillations of the electric field with complex amplitude $\tilde{\vec{E}}_c = i \Omega \tilde{\vec{A}}_c/c - i\vec{K} \tilde{\varphi}_c$ and magnetic field with complex amplitude $\tilde{\vec{B}}_c = i\vec{K} \times \tilde{\vec{A}}_c$. This leads to
\begin{gather}
\bar{H} = H_0 + e^2|\tilde{\vec{A}}_c|^2/(4mc^2\bar{\gamma}^3),\\
\tilde{H} = - e (\vec{P} - e\bar{\vec{A}}/c)\cdot \tilde{\vec{A}}/(mc\bar{\gamma})+ e\tilde{\varphi},
\end{gather}
where $H_0 \doteq mc^2\bar{\gamma}+ e\bar{\varphi}$, $\bar{\gamma} \doteq [1 + (\vec{P} - e\bar{\vec{A}}/c)^2/(mc)^2]^{1/2}$, and quiver terms scaling as second and higher powers of $\tilde{\vec{A}}$ and $\tilde{\varphi}$ are neglected. One can check then that \Eq{eq:Heff} reproduces the OC Hamiltonians derived earlier, and $\Phi \doteq \mc{H} - H_0$ is the well-known ponderomotive potential \cite{foot:pp}.

\msection{Conclusions} In summary, we showed that nonlinear interactions of waves via instantaneous XPM can be cast in the same way as ponderomotive wave-particle interactions in high-frequency EM field. The ponderomotive effect arises when rays of a PW scatter off perturbations of the underlying medium produced by a MW, much like charged particles scatter off a quasiperiodic EM field. The striking parallels with the point-particle dynamics, which itself is generalized by this theory, lead to new methods of wave manipulation, including asymmetric barriers for light.

The work was supported by the NNSA SSAA Program through DOE Research Grant No. DE274-FG52-08NA28553, by the U.S. DOE through Contract No. DE-AC02-09CH11466, and by the U.S. DTRA through Research Grant No. HDTRA1-11-1-0037.


\noindent
\textsl{{Notice:} This manuscript has been authored by Princeton University under Contract Number 
DE-AC02-09CH11466 with the U.S. Department of Energy. The publisher, by accepting the article for publication acknowledges, that the United States Government retains a non-exclusive, paid-up, irrevocable, world-wide license to publish or reproduce the published form of this manuscript, or allow others to do so, for United States Government purposes.}

\end{document}